\documentclass[twocolumn,showpacs,preprintnumbers,amssymb,nofootinbib, aps,prd, eqsecnum ]{revtex4}
\usepackage{graphicx, booktabs, epsf, epsfig}% Include figure files
\usepackage{bm} % Include bold math: \bm{} creates bold letters and symbols in math mode
\usepackage{amsmath}

\def\be{\begin{equation}}
\def\ee{\end{equation}}
\def\beq{\begin{eqnarray}}
\def\eeq{\end{eqnarray}}
\def \bsp{\begin{split}}
\def \ensp{ \end{split} }

\def\o{\omega}
\def\IL{\relax{\rm I\kern-.18em L}}

\def\f{\frac}

\begin{document}

\title{Quasinormal modes of a massless charged scalar field on a small Reissner-Nordstr\"om-anti-de Sitter black hole}

\author{Nami Uchikata}
\email{nami@astr.tohoku.ac.jp}
\author{Shijun Yoshida}
\email{yoshida@astr.tohoku.ac.jp}
 \affiliation{Astronomical Institute, Tohoku University, Aramaki-Aoba, Aoba-ku, Sendai
980-8578, Japan}

\date{\today}

\begin{abstract}
We investigate quasinormal modes of a massless charged scalar field on a small Reissner-Nordstr\"om-anti-de Sitter (RN-AdS) black hole 
both with analytical and numerical approaches. In the analytical approach, by using the small black hole approximation ($r_+ \ll L$), we 
obtain the quasinormal mode frequencies in the limit of $r_+/L\rightarrow 0$, where $r_+$ and $L$ stand for the black hole event horizon radius 
and the AdS scale, respectively. We then show that the small RN-AdS black hole is unstable if its quasinormal modes satisfy the superradiance 
condition and that the instability condition of the RN-AdS black hole in the limit of $r_+/L\rightarrow 0$ is given by $Q>\f{3} {eL}Q_c$, where 
$Q$, $Q_c$, and $e$ are the charge of the black hole, the critical (maximum) charge of the black hole, and the charge of the scalar field, respectively. 
In the numerical approach, we calculate the quasinormal modes for the small RN-AdS black holes with $r_+ \ll L$ and confirm that the RN-AdS 
black hole is unstable if its quasinormal modes satisfy the superradiance condition. Our numerical results show that the RN-AdS black holes with 
$r_+ =0.2L$, $0.1L$, and $0.01L$ become unstable against scalar perturbations with $eL=4$ when the charge of  the black hole 
satisfies $Q \gtrsim 0.8Q_c$, $0.78Q_c$, and $0.76Q_c$, respectively. 
\end{abstract}


\maketitle
\newpage
%%%%%%%%%%%%%%%%%%%%%%%%%%%%%%%%%%%%%%%%%%%%%%%%%%%%%
\section{Introduction}
%%%%%%%%%%%%%%%%%%%%%%%%%%%%%%%%%%%%%%%%%%%%%%%%%%%%%
Superradiance is a phenomenon in which impinging waves are amplified as they scatter off a rotating black hole, e.g., the Kerr black hole, 
by extracting its angular momentum   \cite{zel1, staro, mis}. It is also known that a similar phenomenon, a charge extraction, happens for a charged black hole, 
e.g., the Reissner-Nordstr\"om (RN) black hole or the Kerr-Newman (KN) black hole, if impinging waves have a charge \cite{ bek, hr, rwpu}. 
Using Hawking's area theorem \cite{bek, haw}, Bekenstein has derived the charge 
extraction condition, given by $\o + e \phi<0$, where $\o$ is the frequency of the impinging waves,  $e$ the charge 
of the wave field, and  $\phi $ the electrostatic potential difference between the black hole horizon and the spatial infinity. 
If the effect of rotation is included, the angular momentum and/or charge extraction condition is generalized to $\omega-m\Omega+ e \phi<0$, 
which is  referred to as ``the superradiance condition'' in this study, where $\Omega$ and $m$ denote the rotation angular velocity of the black hole 
and the azimuthal quantum number of the wave field, respectively. Since the quasinormal 
frequencies of the uncharged massless field do not satisfy the superradiance condition, the angular momentum and/or the charge extraction do not occur spontaneously 
in the black hole in asymptotically flat spacetime \cite{leaver, leaver2}. This means that the black hole in asymptotically flat spacetime is stable.

The situation is different for a black hole in  asymptotically anti de Sitter (AdS)  spacetime. The AdS boundary behaves as an effective wall that 
encloses the black hole and waves in AdS space are reflected inward at the AdS boundary. As argued in \cite{cardoso2}, thus, a black hole in AdS 
spacetime may be modeled by a black hole in a spherical container whose length scale is given by the AdS scale $L$ (see Sec. II). 
Thus,  the quasinormal frequencies of the black hole in AdS spacetime are scaled by $L^{-1}$ as long as the black hole event horizon radius $r_+$ 
is sufficiently smaller than the AdS scale $L$, i.e., $r_+ \ll L $. This implies that the nondimensional quasinormal frequency $\omega r_+$ of the black hole 
in AdS spacetime unboundedly decreases with decreasing the nondimensional horizon radius of the black hole $r_+/L$ \cite{cardoso2}.  
For such low frequency normal modes, the superradiance condition will be easily satisfied. (The superradiance condition is given by $\omega r_+-e Q<0$ 
for the RN black hole, where $Q$ is the charge of the black hole.)
Because the superradiant wave amplifications will occur near the black 
hole and will not be affected by the structure at infinity, it is expected that the rotating and/or charged black holes in asymptotically AdS spacetime 
are unstable if the quasinormal mode satisfies the superradiance condition. 
In fact, it has been shown analytically and numerically that a sufficiently small Kerr AdS black hole is unstable and its unstable quasinormal 
frequencies satisfy the superradiance condition \cite{hht, cardoso2, kodama}. 

For the charged AdS black hole in the four dimensional spacetime, there has been no detailed investigation with respect to the relation between the condition 
$\o + e \phi<0$ and the instability of the black hole. 
Most studies on perturbations of the RN-AdS black hole that have been done so far focus on its quasinormal mode only (not on its instability), 
see, e.g., \cite{kono, bk, wla, wma, wlm}, even though Wang et al. \cite{wlm} suggest that the extreme RN-AdS black hole is neutrally stable 
against scalar perturbations. However, there have been at least two exceptions in which the stability of the RN-AdS black hole is argued. 
Abdalla et al. \cite{apop} have studied massive charged scalar perturbations of the RN-AdS black hole by considering the time evolution of the scalar field 
 and have found that RN-AdS black holes with moderate mass and charge become unstable. However, the physical origin of the instability was not discussed in \cite{apop}. 
To confirm the instability, they also tried to obtain the unstable quasinormal modes by solving the standard eigenvalue problem but did not find any 
 unstable quasinormal mode. 

Maeda et al. \cite{mfk} have studied the stability of the RN-AdS black hole around a critical temperature
and have shown that the RN-AdS black hole becomes unstable below a critical temperature. They also found that the charge extraction from the black hole occurs  
as the result of the instability. 
Although the charge extraction happens, they concluded that this instability is not due to superradiance because it is not related to the 
superradiace condition (see Sec IV). 

As for the KN black hole, which is a charged rotating black hole in asymptotically flat spacetime, 
Furuhashi and Nambu have studied the superradiance instability against charged massive scalar field perturbations \cite{fn}. 
They showed that the KN black hole becomes unstable if the quasinormal mode satisfies the superradiance condition. 
They found that the KN black hole becomes unstable due to superradiance for some values of its spin and charge and 
that the superradiance instability does not occur in the RN black hole or in the no-rotation limit.

In this paper, we consider 
massless charged scalar field perturbations of RN-AdS black holes in order to study the relation between instability and the charge extraction due to superradiance. 
We focus on the RN-AdS black holes having a sufficiently small horizon radius, i.e., $r_+ \ll L$. The plan of the paper is the following. In Sec. II we analytically obtain  
the quasinormal modes of the scalar field under the small black hole approximation, $r_+ \ll L$, to derive the instability condition, and 
briefly describe our numerical method. The numerical results are shown in Sec. III. We discuss the relation between instability of the black hole and superradiance 
in Sec. IV. Then the conclusion is in the last section.  
%%%%%%%%%%%%%%%%%%%%%%%%%%%%%%%%%%%%%%%%%%%%%%%%%%%%%%%%%%%
\section{scalar perturbations of RN-AdS black holes} 
%%%%%%%%%%%%%%%%%%%%%%%%%%%%%%%%%%%%%%%%%%%%%%%%%%%%%%%%%%%
%%%%%%%%%%%%%%%%%%%%%%%%%%%%%%%%%%%%%%%%%%%%%%%%%%%%%%%%%%%
\subsection{Reissner-Nordstr\"om AdS black hole}
%%%%%%%%%%%%%%%%%%%%%%%%%%%%%%%%%%%%%%%%%%%%%%%%%%%%%%%%%%%
The metric of the RN-AdS black hole is
\begin{equation}
 ds^2= - \f {\Delta} {r^2} dt^2 +\f {r^2} { \Delta} dr^2 +r^2 (d \theta ^2 + \sin ^2 \theta \, d \phi^2), 
 \label{metric}
\end{equation}
where
\be
\Delta = \f {r^4} {L^2} +r^2 - 2 M r +Q^2 .
\ee
Here $L$ is the AdS scale, defined by $L\equiv\sqrt{- 3 / \Lambda}$ with $\Lambda$ being the cosmological constant . 
The parameters $M$ and $Q$ are the mass and the electrical charge of the black hole, respectively. The line element (\ref{metric}) 
describes the black hole if $\Delta = 0$ has two real roots; the event horizon $r_+$ and the Cauchy horizon $r_-$. The Hawking 
temperature is given by 
\be
T= \f {1} {4 \pi} \left ( \f {d} {dr} \f {\Delta} {r^2} \right )\bigg |_{r=r_+} = \f{ 1 } {4 \pi r_+} \left( 1- \f {Q^2} {r_+^2} + \f {3 r_+^2} {L^2} \right ). 
\ee
Since the charge $Q$ is constrained by the positive definiteness of the Hawking temperature \cite{bk}, one has 
\be
\label{qc}
Q  \leqslant r_+ \sqrt{ 1 + \f {3 r_+ ^2} {L^2}} \equiv Q_c \, .
\ee
For a small black hole, $r_+ \ll L$, the critical charge $Q_c$ becomes $Q_c = r_+(1+O ((r_+/L)^2)) \ll L$. The electromagnetic vector potential 
of the black hole is 
\be
A_\mu dx^\mu = \phi (r) \, dt = (-\f {Q} {r} +C) dt \, ,
\ee 
with an integration constant $C$. For the RN black hole, one takes $C=0$. For the RN-AdS black hole, on the other hand, we often chose 
$C=Q/r_+$ because the potential needs to vanish on the event horizon in the context of the AdS/CFT correspondence (see, for a review, e.g., 
 \cite{herzog, horowitz}). Note that as shown later, 
the integral constant $C$ does not affect the value of Im$(\o)$, but it just displaces the value of Re$(\o)$ as Re$(\o) \to$Re$(\o) + e C$, 
where $\o$ denotes the quasinormal frequency of the black hole.
 
%%%%%%%%%%%%%%%%%%%%%%%%%%%%%%%%%%%%%%%%%%%%%%%%%%%%%%%%%%%
\subsection{Basic equations for scalar perturbations}
%%%%%%%%%%%%%%%%%%%%%%%%%%%%%%%%%%%%%%%%%%%%%%%%%%%%%%%%%%%
We consider massless charged scalar field perturbations of the RN-AdS black hole. The field equation is written as,
\be
(\nabla _{\mu} - i e A_{\mu}) (\nabla ^{\mu} - i e A^{\mu} ) \Phi = 0,
\ee
with $e$ being the charge of the scalar field $\Phi$. 
The master equation is reduced to an ordinary differential equation by  the separation of variables: 
\begin{equation}
\Phi(t,r,\theta,\phi)=e^{-i\omega t } Y _{l m}(\theta, \phi) R(r)\,, \label{separation}
\end{equation}
where $Y_{l m} (\theta, \phi)$ is the spherical harmonic function.  We may assume $\omega$ to be  
${\rm Re}(\omega)>0$ without loss of generality. Thus, in this study, we consider modes with ${\rm Re}(\omega)>0$ only. 
Then we have the radial equation, 
\begin{equation}
 \f {d} {dr} \left ( \Delta \f {d R (r)} {dr} \right )   +\left [ \f {(\o + e \phi) ^2 r^4} {\Delta } -  l (l+1)\right ] R(r)=0,
\label{radial}
\end{equation}
where $l$ is the eigenvalue associated with $Y_{l m} (\theta, \phi)$.
For the numerical calculation, it is useful to rewrite the radial wave equation as 
\be
\f {d ^2 X(r)} {d r_* ^2} + \hat{\o}^2 X (r)= \f {\Delta } {r^2} \left ( \f {r \Delta^{\prime}- 2\Delta } {r^4} + \f {l (l+1)} {r^2} \right ) X(r),
\label{radial2}
\ee
with 
$$ \frac{dr_*}{dr}=\frac{r^2}{\Delta} , \quad X(r) = r R(r),\quad  \hat{\o} = \o + e \phi(r),$$
and the prime denotes $d/dr$. Near the event horizon, the radial wave function has to behave as,
\be
X \sim  e^{- i \hat{\o}
r_*}\,\,,\,\,r\rightarrow r_+\,. \label{bc2} 
\ee
Here we have imposed the condition of no outgoing wave from the event horizon. Near spatial infinity, the radial function is given by 
\be
X \sim A r + \f {B} {r^2}, \quad {\rm as}\ \ r \to \infty,
\label{r-inf}
\ee 
with constants $A$ and $B$. 
To avoid unphysical divergence of the scalar field at infinity, we need to impose $A=0$. Then we have $R \propto r^{-3}$.

%%%%%%%%%%%%%%%%%%%%%%%%%%%%%%%%%%%%%%%%%%%%%%%%%%%%%%%%%%%
\subsection {Analytical condition of instability }
%%%%%%%%%%%%%%%%%%%%%%%%%%%%%%%%%%%%%%%%%%%%%%%%%%%%%%%%%%%
Before doing the numerical calculation, it is useful to have an analytical formula for the instability condition. Following  \cite{cardoso2}, in which  
the instability conditions for the Kerr-AdS black hole and the black hole bomb are obtained with the matched asymptotic expansion, we derive 
the instability condition for the RN-AdS black hole. 
Here, we assume $\o M \ll 1$ in advance. But, this assumption will be justified later.  

In the near region defined by ${\rm Re}(\omega)(r-r_+)\ll 1$, the radial equation is reduced to
\be
 \f {d} {dr} \left ( \Delta \f {d R (r)} {dr} \right )   +\left [ \f {(r_+ -r_ -) ^2 \sigma^2} {\Delta } -  l (l+1)\right ] R(r)=0,
 \label{neq}
\ee
where
\be
\begin{split}
&\sigma  \equiv \f {(\o + e \phi (r_+))  r_+^2} {r_+ -r_-} = \left ( \tilde{ \o} -  e \f {Q} {r_+} \right  ) \f {  r_+^2} {r_+ -r_-}, \\
& \tilde{\o} \equiv \o +e C.
\end{split}
\ee
Introducing a new independent variable and a new radial function, defined by 
\be
z\equiv \f {r-r_+} {r-r_-}, \quad R(r) \equiv z^{i \sigma} (1+z) ^{l+1} F,
\ee
one obtains the hypergeometric equation as
\be
z(1-z) \f {d^2 F} {dz^2} +[\gamma -(\alpha +\beta +1) z ] \f {d F} {dz} - \alpha \beta F = 0,
\ee
with
$$ \alpha = 2 i \sigma+l+1 , \quad  \beta= l+1, \quad \gamma =2 i \sigma +1 .$$ 
Imposing the ingoing wave condition at the event horizon, one has the physically acceptable solution of Eq.\eqref{neq}, given by
\be
R = z^{-i \sigma } F( \alpha - \gamma +1, \beta - \gamma +1; 2-\gamma ;z).
\label{near}
\ee
Here, $F(a,b;c;x)$ is the hypergeometric function. 
To match the solution \eqref{near}  to the far region solution,  one needs an asymptotic form of Eq.\eqref{near} around $|z|=1$, which 
can be obtained by the $z \to 1-z$ transformation law of the hypergeometric function \cite{abramowitz}. We then get
\be
\begin{split}
R & \sim \f {\Gamma [-2 l -1] (r_+ -r_-) ^{l+1}} {\Gamma[-l] \Gamma[-2 i \sigma -l]} r^{-l-1}\\
 &  +\f {\Gamma [2 l +1] (r_+ -r_-) ^{-l}} {\Gamma[l+1] \Gamma[l+1 - 2 i \sigma]} r^ l.
 \label{near2}
\end{split}
\ee

In the far region defined by $r_+ \ll r$,  the radial equation \eqref{radial} becomes
\be
\begin{split}
\left( 1 + \f {r^2} {L^2} \right )  \f {d^2 R} {dr^2} &+ \f {2} {r} \left ( 1+\f {2 r^2} {L^2} \right ) \f  {d R} {dr} \\
& + \left [ \f {\tilde{\o } ^2} {1+r^2/L^2} -\f {l (l+1)} {r^2} \right ] R=0.
\end{split}
\label{far1}
\ee
In this region, the effects of the black hole may be neglected. We have thus  assumed that $M \sim 0$ and $Q \sim 0$. 
 Note that as mentioned, $Q \le Q_c = r_+(1+O ((r_+/L)^2)) \ll L$. Eq.(\ref{far1}) is the same as that of scalar perturbations 
of the pure AdS spacetime if $\tilde{\o }$ is replaced by $\o$. 
Provided one introduces the independent variable and the radial wave function, defined by 
\be
x=1 +\f {r^2} {L^2} , \quad R=x^{\tilde{ \o} L/2} (1-x)^{l/2} F(x)\, ,
\ee
the function $F$ satisfies the hypergeometric equation,
\be
x(1-x) \f {d^2 F} {dx^2} +[\gamma -(\alpha +\beta +1) x ] \f {d F} {dx} - \alpha \beta F = 0,
\label{far}
\ee
where
\be
\alpha = \f {l + 3 +\tilde{ \o} L } {2} , \quad \beta =\f {l+ \tilde{ \o} L } {2},\quad \gamma =1 + \tilde{ \o } L.
\ee
A general solution of Eq.\eqref {far} is given by 
\be
\begin{split}
R& = \tilde{A} x^{-(l+3)/2} (1-x) ^{l/2}F (\alpha, \alpha-\gamma +1; \alpha -\beta +1; 1/x)\\
& +   \tilde{B} x^{-l/2} (1-x) ^{l/2}F (\beta, \beta-\gamma +1; \beta -\alpha +1; 1/x), 
\end{split}
\ee
where $\tilde{A}$ and $ \tilde{B}$ are constants. 
We have to examine the behavior of the far region solution both at spatial infinity and in the intermediate region defined by 
$r \gg r_+$ and ${\rm Re}(\omega)(r-r_+)\ll 1$. 
Near infinity, we have 
\be
R \propto \tilde{A} x ^{-3/2} +  \tilde{B} \quad {\rm as}\ x \to \infty \, .
\ee
The radial function has to behave as $R \sim r^{-3}$ at infinity (see Eq.\eqref{r-inf} with $A=0$). Thus, the boundary condition at infinity is given by $ \tilde{B}=0$.
In the small $\o r$ ($x \to 1$) limit, or in the intermediate region, 
one has 
\be
\begin{split}
R/ \tilde{A} & \sim \f { \Gamma [l+1/2]  r^{-l-1} } {\Gamma[ (3+l) /2 + \tilde{ \o }L /2 ] \Gamma[ (3 + l)/2 - \tilde{ \o } L /2]} \\
 &  + \f { \Gamma [-l-1/2] r^l } {\Gamma[ 1- l /2 - \tilde { \o } L /2 ] \Gamma[ 1 - l/2 + \tilde{ \o } L /2]}  .\label{far2}
\end{split}
\ee
In the limit of $r_+/L \rightarrow 0$, ${\rm Re}(\o) \gg {\rm Im}(\o)$ because waves can hardly leak out of the horizon. In other words, there is no 
black hole effect in the $r_+/L \rightarrow 0$ limit. 
Thus, this radial function must be regular at $r=0$. We then obtain the eigenvalue equation, given by $(3 + l)/2 -\tilde{\o } L/2 = -n+O(r_+/L)$, i.e., 
\be
\tilde {\o}= \f {2n + l +3} {L}\equiv \tilde{\o}_0,
\label{def_w0}
\ee
with an integer $n$ ($n \geqslant 0$) (for a detailed discussion
   of quasinormal modes of the Schwarzschild-AdS black hole in the small
   black hole limit, see \cite{kono2}). If the effect of the black hole is taken into account, the quasinormal frequency $\o_Q$ becomes complex and 
may be written as 
\be
\tilde {\o }_Q \equiv (\o_Q+ e C) = \tilde{\o}_0 + i \delta ,
\ee
for small AdS black holes characterized by $r_+ \ll L$, where the subscript ``{\it Q}''  to the right of $\omega$ is attached to emphasize that the $\omega$ is 
the quasinormal frequency of the black hole. 
Here, $ | \delta | \ll |\mbox{Re} (\tilde{ \o }_0 )| $ has been assumed as already mentioned. 

Matching the near region solution \eqref{near2} with the far region solution \eqref{far2} in the intermediate region,  we obtain the frequency 
correction $\delta$, whose sign basically determines stability of the black hole. By using properties of the gamma function, we finally have  
\be
\begin{split}
\mbox{Im} (\o_Q) 
 = &- \sigma_0 \f {(l !)^2 (l+2+n)! \, 2^{l+3} (2l +1 +2n) !!} {(2l+1) ! (2l)! n! (2l-1) !! (2l+1) !! (2n+3) !!}  \\
& \times  \f {(r_+ -r_-)^{2l+1}} {\pi L^{2l+2}}\prod^ l_ {k=1} (k^2 +\sigma_0 ^2) ,
\label{delta}
\end{split}
\ee
where
\be
\sigma_0  \equiv  \left ( \tilde{ \o}_0 -  e \f {Q} {r_+} \right  ) \f {  r_+^2} {r_+ -r_-} .
\ee
Therefore, if $\sigma_0 < 0$ or 
\be
{\rm Re}(\tilde {\o }_Q )- e \f { Q} {r_+} <0,
\label{insta-con}
\ee
Im$(\o_Q) > 0$ and the wave function grows exponentially with time, which means that the black hole is unstable. Since Re$(\tilde{\o} _Q) = \tilde{\o}_0$, 
this instability condition is independent from $C$. Because the real part of the quasinormal frequencies is given by 
\be
\mbox {Re} (\o_Q) =\f {2n+l+3} {L} - e C, 
\label{w0}
\ee
Re$(\o_Q)$ can vanish at some value of $Q/r_+$ if $C= Q/r_+$ and $e L \gtrsim O (1)$. When Re($\o_Q ) =0$ for the $C=Q/r_+$ case, Im($\o_Q$) also vanishes and the black hole is neutrally stable. 

Introducing the dimensionless charge, defined by $\tilde{Q}\equiv Q/Q_c$, we may, in terms of $\tilde{Q}$ and 
$eL$, rewrite the instability condition (\ref{insta-con}) as 
\be
\tilde{Q} > \f {2n + l +3} {eL} ,
\label{insta-con2s}
\ee
in the small black hole limit. Here, $Q_c=r_+(1+O((r_+/L)^2))$ has been used. This instability condition implies that the lowest order mode, characterized 
by $n=0$ and $l=0$, places the most strict condition on the instability, i.e., $\tilde{Q} > \f{3}{eL}$, at least in the limit of $r_+/L\rightarrow 0$. Since $\tilde{Q} \leqslant 1$, we also see that the small black hole is always stable against the charged scalar perturbations with $e L <3$.

From the behavior of the radial function near the horizon, $R \propto e^{- i \hat{\o} r_*-i\omega t}$, (see Eq.\eqref{bc2}), an observer far from the black hole 
recognizes that the waves are coming out from the black hole if the instability condition is satisfied, 
${\rm Re}(\hat{\o}) = {\rm Re}(\o) +e \phi (r_+) = {\rm Re}(\tilde{\o}) -e Q/r_+ <0$. 
However, the group velocity $-\partial  \hat {\o}/\partial \o$ is always negative, which implies 
that a local observer of the black hole sees only waves going into the horizon, then the physically acceptable boundary condition is satisfied. 
 The same phenomenon occurs in the superradiance of the Kerr black hole. 
%%%%%%%%%%%%%%%%%%%%%%%%%%%%%%%%%%%%%%%%%%%%%%%%%%%%%%%%%%%
\subsection {Numerical method}
%%%%%%%%%%%%%%%%%%%%%%%%%%%%%%%%%%%%%%%%%%%%%%%%%%%%%%%%%%%
As mentioned before, we focus on the small black hole. For the small black hole, the imaginary part of the quasinormal 
frequency is so small that the eigenfunction can be evaluated by direct numerical integration of the wave equation \cite{fn, cardoso2}. 
Therefore, we use the direct numerical integration method to obtain the quasinormal frequency of the black hole. 
Since the behavior of the radial function near infinity is given by $X \propto r^{-2}$,  we may expand $X$ in a power series of $r^{-1}$ as 
\be
X \sim \f {1} {r^2}\left (  b_0 + \f {b_1} {r} +  \f {b_2} {r^2} + \f {b_3} {r^3} +\cdots \right ).
\label{ini2}
\ee 
Near the horizon, the radial function is approximated by
\be
X \sim  e ^{- i \hat{\o}_Q r_*} [a_0 + a_1 (r-r_+) + a_2 (r-r_+)^2 +a_3(r-r_+)^3+\cdots]\,.
\label{ini1}
\ee
Here, the coefficients $a_n$ and $b_n$ have not been explicitly shown because they are obtained straightforwardly. 
The radial equation \eqref{radial2} is integrated outward from a point near the horizon with an initial condition (\ref{ini1}) and inward from a point
at a large radius of $r \sim 1000r_+$  with an initial condition (\ref{ini2}). A Runge-Kutta method is used 
for the numerical integration. Then, we have two solutions at a medium radius $r_m$, i.e. $r _ + < r_m < R \sim 1000r_+$; 
one satisfies Eq.(\ref{bc2}) and the other Eq.(\ref{r-inf}) with $A=0$. When the Wronskian of these two solutions $W (r_m,\o)$ at $r=r_m$ vanishes, 
the two solutions are linearly dependent and simultaneously satisfy the two boundary conditions (\ref{bc2}) and (\ref{r-inf}) with $A=0$. 
To find the eigenvalue $\o_Q$, i.e., the quasinormal modes of the black hole, we solve $W(r_m, \o)=0$ iteratively by a secant method. 
The parameter $r_m$ is varied to check accuracy of the numerical solutions. 
%%%%%%%%%%%%%%%%%%%%%%%%%%%%%%%%%%%%%%%%%%%%%%%%%%%%%%%%%
\section{numerical results}
%%%%%%%%%%%%%%%%%%%%%%%%%%%%%%%%%%%%%%%%%%%%%%%%%%%%%%%%5 
For numerical calculations, 
we normalize all the physical quantities by the AdS scale $L$, e.g., $r_+ \to r_+ /L$, $\o \to \o L$, $Q \to Q/L$ and $e \to e L$. 
We set the electrostatic potential as $\phi(r) = -Q/r$ like the RN black holes' potential. We focus on the fundamental quasinormal 
modes of the small RN-AdS black holes, $r_+\ll 1$, because the most profound unstable mode usually appears in the fundamental mode. 
We calculate sequences of the quasinormal mode increasing the value of $Q$ from $Q=0$ with 
a fixed horizon radius and charge of the scalar field. The quasinormal frequency of the Schwarzschild AdS (SAdS) black hole 
is used as the initial guess for seeking the mode. 
When showing the results, we use the dimensionless charge $\tilde{ Q}$, defined by $\tilde{ Q}\equiv Q/ Q_c $. Thus, we have $0 \leqslant\tilde{ Q} \leqslant 1$.  

For the $l=0$ fundamental modes, we display representative values of the complex frequencies of the black holes with $r_+ = 0.2$, $0.1$, $0.01$, and 
$0.001$ in TABLEs I-IV. 
In TABLE IV, we also show representative values given by the analytical formula \eqref{delta} for the black holes with $r_+=0.01$ and $0.001$.  We exhibit the real 
and imaginary parts of the frequencies as functions of $\tilde{ Q}$ in FIGs. 1-11. As the value of the charge of the black hole increases, our numerical 
method does not work well due to some numerical difficulties, thus we show reliable results with sufficient accuracy only  in the tables and the figures.  

For the black hole with $r_+=0.2$, we show the $l=0$ quasinormal mode frequencies of the scalar fields having $e=0$, $2$, and $4$ 
in TABLE I and  FIGs.~1 and 2. 
We use the frequency of the SAdS black hole given in \cite{cardoso} as the initial guess for seeking the modes.  
We can obtain reliable results in the range $0 \leqslant \tilde{Q} \leqslant 0.85$ for the real part of the frequency and $0\leqslant \tilde{Q} \leqslant 0.8$ for 
the imaginary part of the frequency. 
For the $e=4$ case, as shown in FIG.~2, the imaginary part of the frequency increases as the charge of the black hole increases.  
We then see that Im$(\o) > 0$ for $\tilde{Q} \geqslant 0.8$. In this range of the black hole charge, the frequency satisfies Re$(\o) - e Q/r_+<0$. 
For the $e=0$ case, on the other hand, the real and imaginary parts of the frequency decrease as the black hole charge increases. 
As for the $e=2$ case, the both Re$(\o)$ and Im$(\o)$ change mildly as the black hole charge increases. 

For the black holes with $r_+ = 0.1$ and $0.01$, we use the normal mode frequencies of the pure AdS spacetime, i.e., $\o = 2n+l+3$, as the initial 
guesses for seeking the modes.  For the case of $r_+=0.1$ and $e=Q=0$, Re($\o$) and Im($\o$) we obtain are in a good agreement with the results given 
in \cite{kono2}. The results for the $l=0$ modes having $e=0$, $2$, and $4$ are displayed in TABLEs II and III, and FIGs.~3-6. We obtain reliable 
results in the range $0\leqslant \tilde{Q} \leqslant 0.95$ 
for both the real and imaginary parts of the frequencies.  We see that the results for the black holes with  $r_+ = 0.1$ and $0.01$ are similar to those for 
the black hole with  $r_+=0.2$. The oscillation frequencies increase as the black hole charge increases, when the charge of the scalar field is $e=4$, while 
the oscillation frequencies decrease as the black hole charge increases, 
when the scalar field is uncharged. It is found that for the scalar perturbations with $e=4$, Im$(\o) >0$ if $\tilde {Q } \geqslant 0.78$  for the $r_+=0.1$ black hole 
and  if $\tilde {Q } \geqslant 0.76$ for the $r_+=0.01$ black hole. Again, we certainly have Re$(\o)-e Q/r_+<0$ 
for the modes characterized by Im$(\o) >0$ or the unstable modes. 

We also calculate the $l=1$ fundamental modes having $e=0$, $2$, and $4$ for the black hole with $r_+ =0.01$ and the results are given in FIGs.~7 and 8. 
We see that although the qualitative properties of the real parts of the frequency are basically independent of the value of $l$, 
the qualitative properties of the imaginary parts of the frequency for the $l=1$ modes are different from those for the $l=0$ modes . 
It is found that for the $l=1$ perturbations, regardless of the value of $e$, the imaginary parts of the frequency increase as the black hole charge 
increases and approach to zero although they stay negative. 

Now let us compare the instability conditions obtained by the numerical calculations with the analytical instability condition (\ref{insta-con2s}). 
For the $l=0$ perturbations with $eL=4$, we have the instability condition $\tilde {Q} > 3/4= 0.75$ from Eq.(\ref{insta-con2s}). We see that this gives us an accurate 
approximation formula for the stability criterion of the black hole. Even for the black hole with $r_+/L=0.2$, which is the largest among the black holes considered in 
the present study, the relative error included in Eq.(\ref{insta-con2s}) is less than $7$ percent. As for the $l=1$ perturbations with $eL=4$, we have the  
analytical instability condition, $\tilde{Q}>1$. This means that there is no unstable mode associated with $l=1$ and $eL=4$, which is indeed 
consistent with our numerical results. We also observe that there is no unstable mode for the scalar
perturbations having $eL<3$ as argued in Sec. II.c.

To see the effect of the integral constant $C$ on the quasinormal frequency,  in FIG.~9, we show the results for the black hole with $r_+=0.01$ in the 
case of $C=Q / r_+$ or $\phi(r) = Q/r-Q/ r_+$.  We observe the following properties. 
As expected from Eq.\eqref{w0}, Re($\o_Q$) is approximated well by $\mbox {Re} (\o_Q) = \f {2n+l+3} {L} - e {Q\over r_+} $ because 
the small black hole is considered. When the scalar field has 
a positive charge, the oscillation frequencies linearly decrease as the charge of the black hole increases. Note that for the $e=0$ case, the result is the same as that of 
the $C=0$ case. For all the imaginary parts of the frequency we calculate, we obtain exactly the same result as the result of the $C=0$ case. 

In order to check our numerical code, 
we  compare our numerical results with the analytical solution \eqref {delta} obtained in the present study. We show the results for the $e=4$ scalar 
perturbations of the black holes with $r_+ =0.01$ and $0.001$ in  TABLE IV and FIGs. 10 and 11. These show that our numerical results are 
in a good agreement with the analytical ones as long as the horizon radius of the black hole is sufficiently small.  

For all the results obtained in the present study, the real parts of the quasinormal mode frequency for the $e=0$ perturbations are of monotonically increasing 
functions of the black hole charge. This makes a sharp contrast with that for the large RN-AdS black hole cases \cite{bk}.
%%%%%%%%%%%%%%%%%%%%%%%%%%%%%%%%%%%%%%%%%%%%%%%%%%%%%
\section{instability and superradiance}
%%%%%%%%%%%%%%%%%%%%%%%%%%%%%%%%%%%%%%%%%%%%%%%%%%%%%
As mentioned in Introduction, 
it is expected that a sufficiently small Reissner-Nordstr\"om AdS black hole is unstable against the charged scalar perturbations due to 
the superradiance charge extraction. We have proved that this expectation is correct by calculating 
the quasinormal modes of a massless charged scalar field on the Reissner-Nordstr\"om AdS black hole. 
We have confirmed both with the analytical and numerical approaches that the small RN-AdS black hole becomes unstable 
if the normal mode satisfies the superradiance condition. The studies on the stability of the black hole and the superradiance 
that have been done so far suggest that the superradiant condition indeed plays an important roll in stability of the black hole 
irrespective of the structure of infinity (see, e.g.,  \cite{staro, rwpu, cardoso2,hht,kodama,fn}). 

For a small RN-AdS black hole, $r_+ \ll L$, we have Re$(\o)L =O(1)$ (see Eq. (\ref{def_w0})). Thus, the superradiance condition can be satisfied 
for a scalar field having $eL = O(1)$ if the black hole has a moderate charge, $Q/r_+=O(1)$.  This can be confirmed in the present results (see Sections II.C and III). 
For a large RN-AdS black hole, $r_+ \gg L$, on the other hand, the superradiance condition can be satisfied for a scalar field with a sufficiently large charge,
$eL \gg 1$, because the quasinormal mode  frequency is characterized by Re$(\o) L \gg 1$  \cite{bk, wla, wma, hohu}. 
But, so far, we have not known whether or not a large RN-AdS black hole becomes unstable due to the superradiance charge extraction. 

Maeda et al. \cite{mfk} have shown that the RN-AdS black holes having a large $r_+$ become unstable below a critical temperature and 
that the instability results in the charge extraction from the black hole. However, the instability they found is independent of the sign of the scalar field charge $e$. 
Thus they concluded that the instability is not directly related with the superradiance. Note that the instability studied  in this paper depends on the sign of $e$, i.e., 
the instability condition \eqref{insta-con} is never satisfied for the black hole with $Q>0$ if $e<0$. Therefore, it is necessary to investigate the stability of a large 
RN-AdS black hole in detail and to see whether or not there exists the quasinormal modes whose Im$(\o)$ is proportional to $\o+e \phi$. 

Not only the AdS boundary but also the mass of  the perturbation field behaves like an effective confining wall around a black hole. Furuhashi and Nambu \cite{fn} have 
investigated charged massive scalar perturbations of the Kerr-Newman black hole, which is a rotating charged black hole in asymptotically flat spacetime. 
They demonstrated that the instability condition is given by the superradiance condition and that instability appears only for the rotating black hole and 
there is no instability in the RN black hole. This means that the confining wall created by the mass of the scalar field cannot make the charged spherically 
symmetric black hole unstable. This is quite a contrast to the case of the  AdS boundary examined in the present study. 
%%%%%%%%%%%%%%%%%%%%%%%%%%%%%%%%%%%%%%%%%%%%%%%%%%%%%
\section{Conclusions}
%%%%%%%%%%%%%%%%%%%%%%%%%%%%%%%%%%%%%%%%%%%%%%%%%%%%%
We have investigated the quasinormal modes of the massless charged scalar field on the RN-AdS black holes both analytically and 
numerically for the cases of $r_+ \ll L$. In the analytical approach, by using the small black hole approximation ($r_+ \ll L$), 
we obtain the quasinormal mode frequencies in the limit of $r_+/L\rightarrow 0$.  We then show that the small RN-AdS black holes are unstable 
if their quasinormal modes satisfy the superradiance condition and that the instability condition of the RN-AdS black hole in the limit of $r_+/L\rightarrow 0$ 
is given by $Q>\f{2n + l +3} {eL}Q_c$. We also calculate the quasinormal modes numerically for the small RN-AdS black holes 
with $r_+ \ll L$ and confirm that the imaginary part of the frequency certainly changes the sign when the superradiance condition is satisfied. 
Our numerical results show that the black holes with $r_+ =0.2L$, $0.1L$, and $0.01L$ becomes unstable against  scalar perturbations with $eL=4$ when 
the charge of  the black hole satisfies $\tilde{Q} \gtrsim 0.8$, $0.78$, and $0.76$, respectively. 
We compare our numerical results with the analytical results and confirm that our numerical results are perfectly consistent with the analytical ones.

As mentioned before, it is interesting and important to generalize the present analysis to the large RN-AdS black hole cases and to investigate the relation between 
the superradiance condition and the instability. 
This remains for future work. 
%%%%%%%%%%%%%%%%%%%%%%%%%%%%%%%%%%%%%%%%%%%%%%%%%%%%%%%%%%%
\section*{Acknowledgements}
%%%%%%%%%%%%%%%%%%%%%%%%%%%%%%%%%%%%%%%%%%%%%%%%%%%%%%%%%%%
We would like to thank T. Futamase for discussions. N.U. is supported by the GCOE Program ``Weaving Science Web beyond 
Particle-matter Hierarchy'' at Tohoku University.
%%%%%%%%%%%%%%%%%%%%%%%%%%%%%%%%%%%%%%%%%%%%%%%%%%%%%%%%%%%

\clearpage
\begin{table}
\caption{The frequencies of the $l=0$ fundamental modes for the black hole
with $r_+=0.2$. The critical charge of this black hole is $Q_c = 0.21166$.}
\begin{ruledtabular}
\begin{tabular}{ l l l l}\hline
$e$&$Q/Q_c $& Re($\omega$) & Im($\omega$) 
\\ \midrule
0&0\footnote{The results of Re$(\omega)$ and Im($\omega$) of $Q=0$ are from \cite{cardoso}. }&2.4751&-3.8992 $\times 10^{-1}$\\
&0.2&2.4542  & -4.0189 $\times 10^{-1}$\\
&0.4&2.3910 & -4.4309 $\times 10^{-1}$\\
&0.6& 2.2887 & -5.3421 $\times 10^{-1}$ \\ 
&0.8 &2.1920 &  -6.9796 $\times 10^{-1}$\\
  \midrule
2&0.2&   2.5925 &  -3.2028 $\times 10^{-1}$\\
&0.4&2.6707 &-2.7170 $\times 10^{-1}$\\
&0.6 & 2.7050 & -2.4025 $\times 10^{-1}$ \\
 &0.8&2.6809& -2.3652 $\times 10^{-1}$ \\
\midrule
4&0.2& 2.7340    &  -2.4768 $\times 10^{-1}$\\
&0.4&2.9646&-1.4129 $\times 10^{-1}$\\
&0.6 & 3.1707& -5.8312 $\times 10^{-2}$ \\
 &0.8&3.3616& 1.5556 $\times 10^{-3}$ 
\\ \bottomrule
\end{tabular}
\end{ruledtabular}
\end{table}

\begin{table}
\caption{The frequencies of the $l=0$ fundamental modes for the black hole
with $r_+=0.1$. The critical charge of this black hole is $Q_c = 0.10148$.}
\begin{ruledtabular}
\begin{tabular}{ l l l l}\hline
$e$&$Q/Q_c $& Re($\omega$) & Im($\omega$) 
\\ \midrule
0&0 &2.6928&-1.0095 $\times 10^{-1}$\\
&0.2&2.6801  &  -1.0434 $\times 10^{-1}$\\
&0.4&  2.6410 &  -1.1625  $\times 10^{-1}$\\
&0.6&2.5723   &  -1.4417 $\times 10^{-1}$ \\ 
&0.8 &2.4787  &  -2.0756  $\times 10^{-1}$\\
&0.9& 2.4332 & -2.5493 $\times 10^{-1}$\\
  \midrule
2&0.2&    2.7614 &  -8.1233 $\times 10^{-2}$\\
&0.4& 2.8058 & -6.7765 $\times 10^{-2}$\\
&0.6 & 2.8256 &  -5.8950 $\times 10^{-2}$ \\
 &0.8& 2.8161 & -5.6894 $\times 10^{-2}$ \\
&0.9& 2.7977 &-6.3772 $\times 10^{-2}$\\
\midrule
4&0.2& 2.8414    &  -6.1698 $\times 10^{-2}$\\
&0.4&2.9650 &-3.4408 $\times 10^{-2}$\\
&0.6 &3.0672  &-1.3945 $\times 10^{-2}$ \\
 &0.8&3.1515&  1.7314 $\times 10^{-3}$ \\
&0.9& 3.1878 &  6.5742 $\times 10^{-3}$
\\ \bottomrule
\end{tabular}
\end{ruledtabular}
\end{table}

\begin{table}
\caption{The frequencies of the $l=0$ fundamental modes for the black hole
with $r_+=0.01$. The critical charge of this black hole is $Q_c=1.0001 \times 10^{-2}$.}
\begin{ruledtabular}
\begin{tabular}{ l l l l}\hline
$e$&$Q/Q_c $& Re($\omega$) & Im($\omega$) 
\\ \midrule
0&0 &2.9737&-5.4711 $\times 10^{-4}$\\
&0.2&2.9727  &  -5.4887 $\times 10^{-4}$\\
&0.4&  2.9695 &-5.5452 $\times 10^{-4}$\\
&0.6&2.9642   &  -5.6548 $\times 10^{-4}$ \\ 
&0.8 &2.9566  &  -5.8781  $\times 10^{-4}$\\
&0.9&  2.9519 & -6.1510 $\times 10^{-4}$\\
  \midrule
2&0.2&   2.9797   &  -4.6984 $\times 10^{-4}$\\
&0.4& 2.9835 & -3.9639 $\times 10^{-4}$\\
&0.6 & 2.9853 & -3.2537 $\times 10^{-4}$ \\
 &0.8&2.9849 & -2.5593 $\times 10^{-4}$ \\
&0.9& 2.9839 &-2.2232 $\times 10^{-4}$\\
\midrule
4&0.2& 2.9866    &  -3.9296 $\times 10^{-4}$\\
&0.4&2.9974 &-2.4694 $\times 10^{-4}$\\
&0.6 &3.0061  &-1.0589 $\times 10^{-4}$ \\
 &0.8&3.0127& 3.2603 $\times 10^{-5}$ \\
&0.9& 3.0153 &  1.0128 $\times 10^{-4}$
\\ \bottomrule
\end{tabular}
\end{ruledtabular}
\end{table}

\begin{table}
\caption{The frequencies of the $l=0$ fundamental modes for the black hole
with $e=4$.}
\begin{ruledtabular}
\begin{tabular}{ l l l l}\hline
$r_+$&$Q/Q_c $& Im($\omega$) (numerical) & Im($\omega$) (analytical)
\\ \midrule
0.01&0 &-5.4711 $\times 10^{-4}$&-5.0484 $\times 10^{-4}$\\
&0.2&-3.9296 $\times 10^{-4}$ &  -3.7120 $\times 10^{-4}$\\
&0.4& -2.4694 $\times 10^{-4}$&-2.3719 $\times 10^{-4}$\\
&0.6&-1.0589 $\times 10^{-4}$  & -1.0283 $\times 10^{-4}$ \\ 
&0.8 &3.2603 $\times 10^{-5}$ &  3.1865  $\times 10^{-5}$\\
&0.9&  1.0128 $\times 10^{-4}$& 9.9347 $\times 10^{-5}$\\
  \midrule
  0.001&0 &-5.1245 $\times 10^{-6}$&-5.0886 $\times 10^{-6}$\\
 &0.2&   -3.7516  $\times 10^{-6}$ &  -3.7326 $\times 10^{-6}$\\
&0.4&-2.3850 $\times 10^{-6}$ &-2.3763 $\times 10^{-6}$\\
&0.6 & -1.0225 $\times 10^{-6}$ & -1.0196 $\times 10^{-6}$ \\
 &0.8&3.3825 $\times 10^{-7}$ & 3.3740 $\times 10^{-7}$ \\
&0.9& 1.0186 $\times 10^{-6}$ &1.0160 $\times 10^{-6}$
\\ \bottomrule
\end{tabular}
\end{ruledtabular}
\end{table}

\begin{figure}
 \includegraphics [height =5.8cm, width=7.5cm] {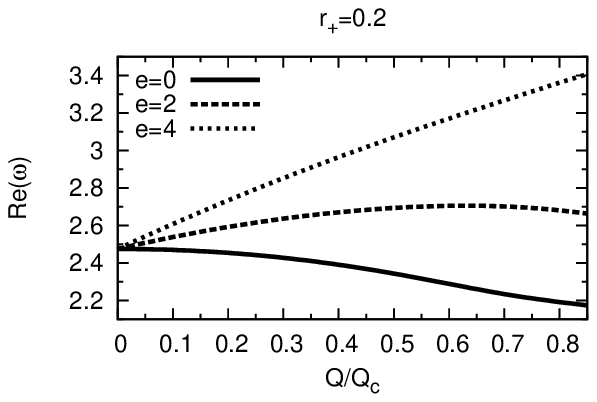}
 \caption{Real parts of $\omega$ of the $l=0$ fundamental modes, given as functions of $\tilde{Q}$,  for the black hole with $r_+=0.1$. }
 \label{rp1.0-l=0,1}
 \end{figure}

\begin{figure}
 \includegraphics  [height =5.8cm, width=7.5cm] {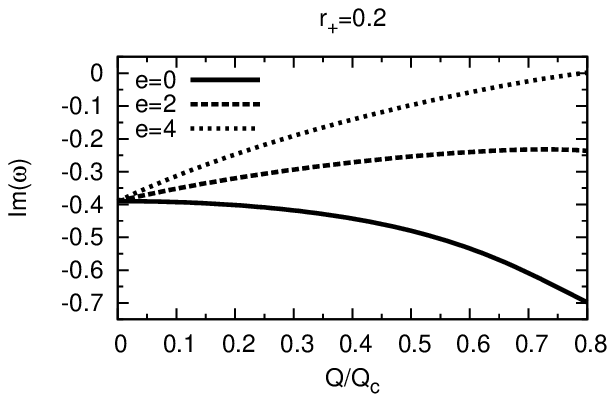}
 \caption{Imaginary parts of $\omega$ of the $l=0$ fundamental modes, given as functions of $\tilde{Q}$,  for the black hole with $r_+=0.2$. }
 \label{rp1.0-l=0,1}
 \end{figure}

\begin{figure}
 \includegraphics  [height =5.8cm, width=7.5cm] {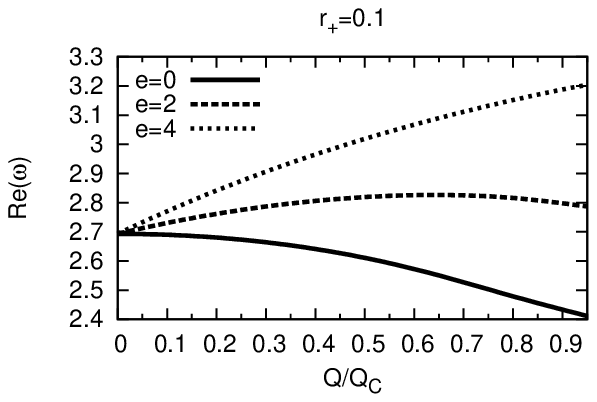}
 \caption{Real parts of $\omega$ of the $l=0$ fundamental modes, given as functions of $\tilde{Q}$,  for the black hole with $r_+=0.1$. }
 \label{rp1.0-l=0,1}
 \end{figure}
 
 \begin{figure}
 \includegraphics  [height =5.8cm, width=7.5cm] {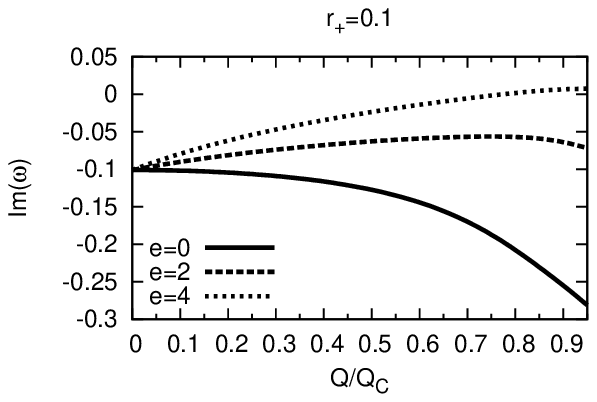}
 \caption{Imaginary parts of $\omega$ of the $l=0$ fundamental modes, given as functions of $\tilde{Q}$,  for the black hole with $r_+=0.1$. }
 \label{rp1.0-l=0,1}
 \end{figure}

\begin{figure}
 \includegraphics  [height =5.8cm, width=7.5cm] {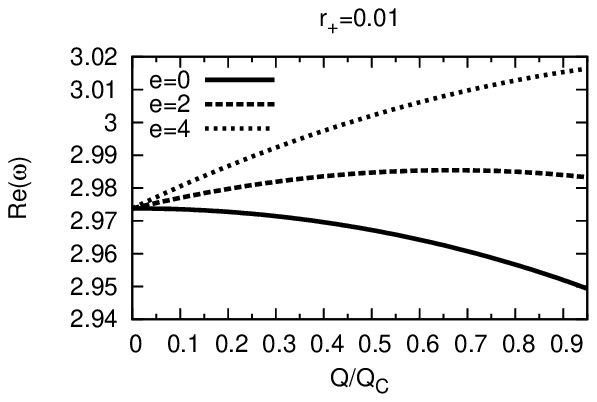}
 \caption{Real parts of $\omega$ of the $l=0$ fundamental modes, given as functions of $\tilde{Q}$,  for the black hole with $r_+=0.01$. }
 \label{rp1.0-l=0,1}
 \end{figure}
 
 \begin{figure}
 \includegraphics  [height =5.8cm, width=7.5cm] {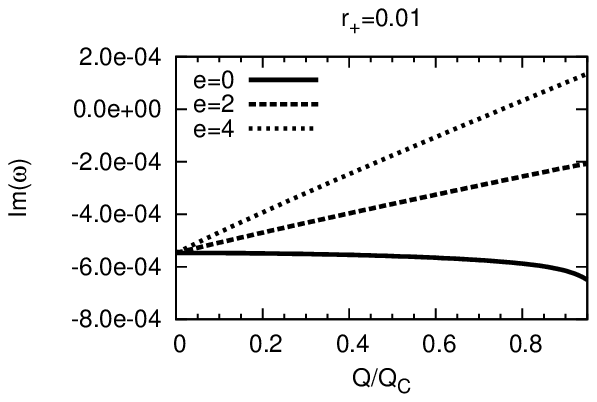}
 \caption{Imaginary parts of $\omega$ of the $l=0$ fundamental modes, given as functions of $\tilde{Q}$,  for the black hole with $r_+=0.01$. }
 \label{rp1.0-l=0,1}
 \end{figure}
 
 \begin{figure}
 \includegraphics  [height =5.8cm, width=7.5cm] {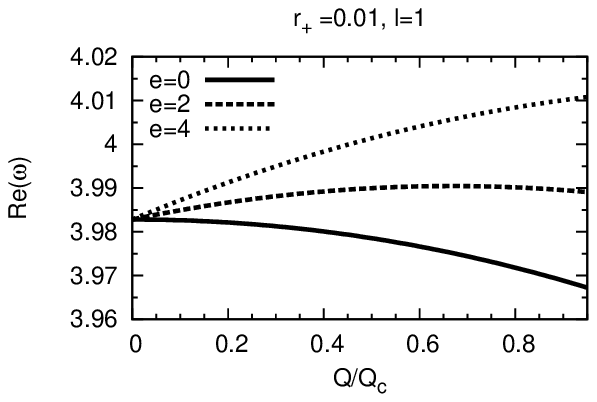}
 \caption{Real parts of $\omega$ of the $l=1$ fundamental modes, given as functions of $\tilde{Q}$,  for the black hole with $r_+=0.01$. }
 \label{rp1.0-l=0,1}
 \end{figure}
 
 \begin{figure}
 \includegraphics  [height =5.8cm, width=7.5cm] {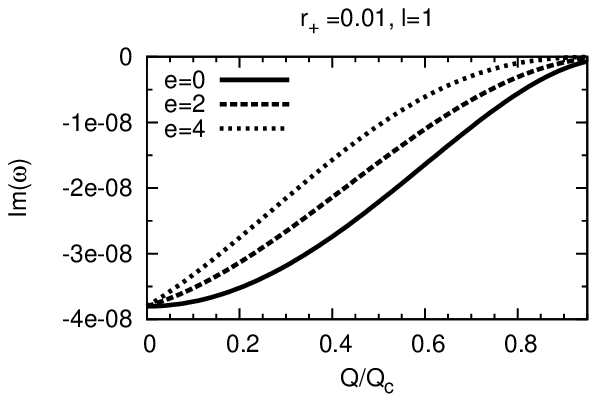}
 \caption{Imaginary parts of $\omega$ of the $l=1$ fundamental modes, given as functions of $\tilde{Q}$,  for the black hole with $r_+=0.01$. }
 \label{rp1.0-l=0,1}
 \end{figure}

\begin{figure}
 \includegraphics  [height =5.8cm, width=7.5cm] {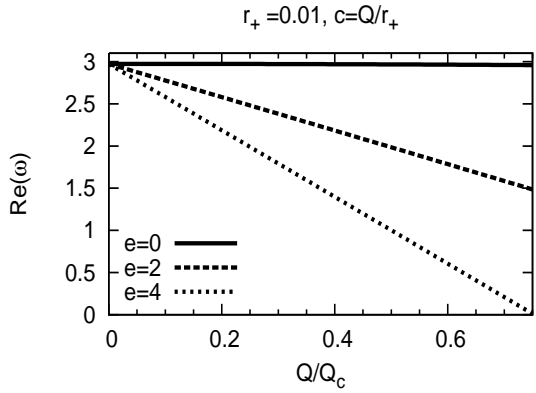}
 \caption{Real parts of $\omega$ of the $l=0$ fundamental modes, given as functions of $\tilde{Q}$,  for the black hole with $r_+=0.01$ having an electrostatic potential $\phi(r) = Q/r_+ -Q/r$. }
 \label{rp1.0-l=0,1}
 \end{figure}

\begin{figure}
 \includegraphics  [height =5.8cm, width=7.5cm] {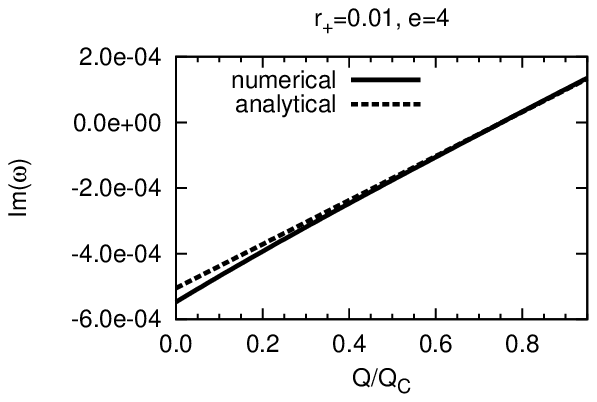}
 \caption{Imaginary parts of $\omega$ of the $l=0$ fundamental modes, given as functions of $\tilde{Q}$,  for the black hole with $r_+=0.01$ in a scalar field having $e=4$. }
 \label{rp1.0-l=0,1}
 \end{figure}
 
\begin{figure}
 \includegraphics  [height =5.8cm, width=7.5cm] {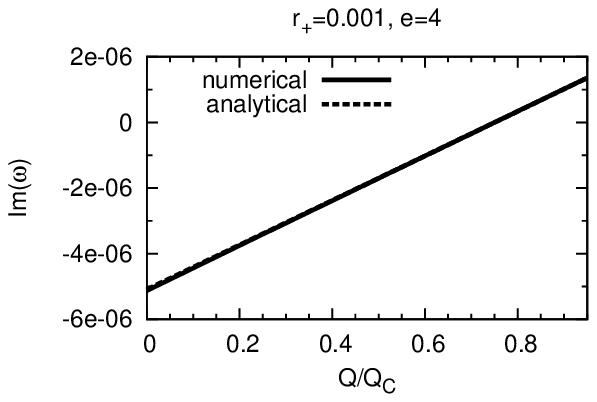}
 \caption{Imaginary parts of $\omega$ of the $l=0$ fundamental modes, given as functions of $\tilde{Q}$,  for the black hole with $r_+=0.001$ in a scalar field having $e=4$. }
 \label{rp1.0-l=0,1}
 \end{figure}

\end{document}